\documentclass[prb,twocolumn,showpacs,preprintnumbers,amssymb]{revtex4}
\usepackage{graphicx}% Include figure files
\usepackage[tbtags]{amsmath}
\usepackage{bm}% bold math

\begin{document}

\title
{Low frequency Rabi spectroscopy of dissipative two level systems.
The dressed state approach. }
\author{Ya. S. Greenberg}
\affiliation{Novosibirsk State Technical University, 20 K. Marx
Ave., 630092 Novosibirsk, Russia.}

\date{\today}
\begin{abstract} We have analyzed a dissipative
two level quantum system (TLS) which
is continuously and simultaneously irradiated by a high and low frequency excitation.
The interaction of the TLS with a high frequency excitation
is considered in the frame of the dressed state approach.  A linear
response of the coupled TLS and corresponding photon field system to a signal whose
frequency is of the order of the Rabi frequency is found. The
response exhibits undamped low frequency oscillations, whose
amplitude has a clear resonance at the Rabi frequency with the
width being dependent on the damping rates of the TLS. The
method can be useful for low-frequency Rabi spectroscopy in
various physical systems described by a two-level Hamiltonian,
such as nuclei spins in NMR, double well quantum dots,
superconducting flux and charge qubits, etc. The application of
the method to a superconducting flux qubit and to the detection of
NMR is considered in detail.

\end{abstract}

\maketitle
\section{Introduction}
It is well known that under resonant irradiation a quantum
two-level system (TLS) can undergo coherent (Rabi) oscillations.
The frequency of these oscillations is proportional to the
amplitude of the resonant field \cite{Rabi} and is much lower than
the gap frequency of the TLS. This effect is widely used in molecular
beam spectroscopy \cite{beam}, and in quantum optics
\cite{Raimond}. During the last few years it has been prove
experimentally that Rabi spectroscopy can serve as a valuable tool
for the determination of relaxation times in solid-state quantum
mechanical two-level systems, qubits, which are to be used for quantum
information processing \cite{Makh}. Normally, these systems  are
strongly coupled to the environment, which results in the fast
damping of the Rabi oscillations. It prevents the use of conventional
continuous measurements schemes for their detection, although
special schemes for the detection of coherent oscillations through
a weak continuous measurement of a TLS were proposed in
\cite{Averin,Kor1,Kor2}. Consequently, Rabi oscillations are
measured by using a pulse technique which harnesses the statistics of
switching events of the occupation probability between two energy
levels with the excitation and readout being taken at the gap
frequency of the TLS, which normally lies in the GHz range
\cite{Nakamura,Vion,Martinis,Chiorescu}.

Recent successful development of the method of low frequency
characterization of flux qubits by weak continuous
measurements in the radio frequency domain (see review paper
\cite{Ilich1} and references therein) allowed the first
spectroscopic monitoring of Rabi oscillations with a
low-frequency tank circuit which had been tuned to the Rabi
frequency of the flux qubit \cite{Ilich2}. These experiments
stimulated theorists to study different methods for the
detection of Rabi oscillations involving low frequency
(compared to the energy gap between the two levels) electronic
circuitry \cite{Smirnov1,Smirnov2,Hauss,Hauss1}.

One of the methods for the detection of Rabi resonance at low
frequencies has been suggested in papers \cite{Green1,Green2}. The
method involves irradiating a TLS continuously and simultaneously by two external
sources. The first source with frequency $\omega_0$, which is
close to the energy gap between the two levels, excites the
low-frequency Rabi oscillations. Normally, Rabi oscillations are
damped out with a rate, which depends on how strongly the system
is coupled to the environment. However, if a second low-frequency
source is applied simultaneously to the TLS it responds with
persistent low frequency oscillations. The amplitude of these
low-frequency oscillations has a resonance at the Rabi frequency
and its width is dependent on the damping rates of the system.
In papers \cite{Green1,Green2} we analyzed the effect in the frame
of Bloch equations for the quantities $\langle\sigma_Z\rangle$,
$\langle\sigma_Y\rangle$, $\langle\sigma_X\rangle$, where the brackets
denote the averaging over the environmental degrees of freedom. Two
external sources at high and low frequency were incorporated in
the structure of Bloch equations from the very beginning. We
showed analytically as well as by direct computer simulations of
the Bloch equations that the quantities $\langle\sigma_Z\rangle$,
$\langle\sigma_Y\rangle$, $\langle\sigma_X\rangle$ exhibit
undamped oscillations with a resonance at the Rabi frequency.

The present paper differs from \cite{Green1,Green2} in that here we
study the problem within a dressed state approach, which is well
known in quantum optics \cite{Coen}. This approach has a great advantage over the method used in \cite{Green2}. In that paper the effect appeared purely mathematically as the solution of the nonlinear Bloch equations for the averaged quantities, which are not quantum in their nature. It was not clear whether the effect had a quantum nature or it was a consequence of nonlinear structure of Bloch equations.

By using the dressed state approach we show here that the effect  has definitely a quantum nature.
From the Heisenberg quantum equations of motion we derive the Bloch like equations for the elements of
the reduced density matrix and find the low frequency susceptibilities
of the TLS coupled to a photon field. We show that, as distinct from the cases in quantum optics, in dissipative solid state TLS there
exists an interaction which can induce transitions between
the dressed Rabi levels. These transitions result in the low
frequency response of the system, with the resonance being at the
Rabi frequency.
In addition, the dressed state approach allowed us to find important contribution to slow dynamics which comes from the transitions between the levels which are spaced apart by the gap frequency $\omega_0$. This contribution is described by the low frequency dynamics of $\kappa$ matrix elements (see below), and it cannot be obtained within the approach used in \cite{Green2}.
Another advantage of the dressed state approach is that the expressions we obtain for low frequency susceptibility for two level system have a rather general nature and can be applied for the investigation of specific physical systems.
As an illustration of our approach in the paper we study in detail two rather different physical systems: the flux qubit and real spin under NMR conditions.
Our approach allows us also to obtain some interesting results which could not be obtained within the method used in \cite{Green2}. For example, we show that the population of Rabi levels becomes inverted under high frequency irradiation, that can lead to the heating of the low frequency source as was shown in Section 7A for the resonant tank of the flux qubit.

The paper is organized as follows. In Section II, for the sake of
the completeness we give a brief overview of a TLS interacting with
a one mode laser field which is tuned to the gap of the TLS. The
structure of the energy levels of the global system (TLS and associated laser field)
consists of manifolds where the spacing between the two levels in a given
manifold is equal to the Rabi frequency. We write down the wave
functions for these two levels and calculate the transition
amplitudes between them which result from the low frequency
excitation. In Section III we define the density matrix in
an uncoupled basis and write down the phenomenological rate equations
for the elements of the density matrix.
In Section IV the elements
of the reduced density matrix (the density matrix traced over the
photon number $N$) are defined, and the Bloch like rate equations for the
elements of the reduced density matrix are derived.
For the case of small high frequency detuning the steady state
solutions to these equations are found. It is shown that under
high frequency irradiation the population of the Rabi levels
becomes inverted. In Section V the Bloch equations for the reduced
density matrix are modified to include the low frequency
excitation. The linear low frequency susceptibilities for the
response of the coupled TLS and associated photon field system to a low frequency
excitation are found in Section VI both for arbitrary and small
high frequency detuning. Two important applications of the method
(superconducting flux qubit and NMR spins) are considered in
detail in Section VII.
\section{Interaction of a TLS with a laser field in the frame of dressed states. A brief overview}
We start with the Hamiltonian of a TLS subjected to a
high frequency field:
\begin{equation}\label{Ham1}
H = \frac{\Delta }{2}\sigma _x  + \frac{\varepsilon }{2}\sigma _z
+\hbar\omega_0(a^+a+1/2)+H_{int}
\end{equation}

Here the first two terms describe an isolated TLS, which can be used to model
a great variety of situations in physics and chemistry: from a
spin-(1/2) particle in a magnetic field to superconducting flux
and charge qubits \cite{Makh, Grif}. In order to be exact we
use the TLS in (\ref{Ham1}) to describe a double-well system
where only the ground states of the two wells are occupied, with
$\Delta$ being the energy splitting of a symmetric ($\varepsilon =
0$) TLS due to quantum tunnelling between two wells. The quantity
$\varepsilon$ is the bias, the external energy parameter which
makes the system asymmetric. The third term in (\ref{Ham1}) is the
Hamiltonian of the laser mode, $a^+$ and $a$ being the creation and
annihilation operators. The last term in (\ref{Ham1}) describes
the interaction of the TLS with a laser field. This interaction
modulates the energy asymmetry between the two wells:

\begin{equation}\label{Ham2}
   H_{int} = - \frac{1}{2}\sigma _z F(a^++a)
\end{equation}

\begin{figure}
  \includegraphics[width=7 cm]{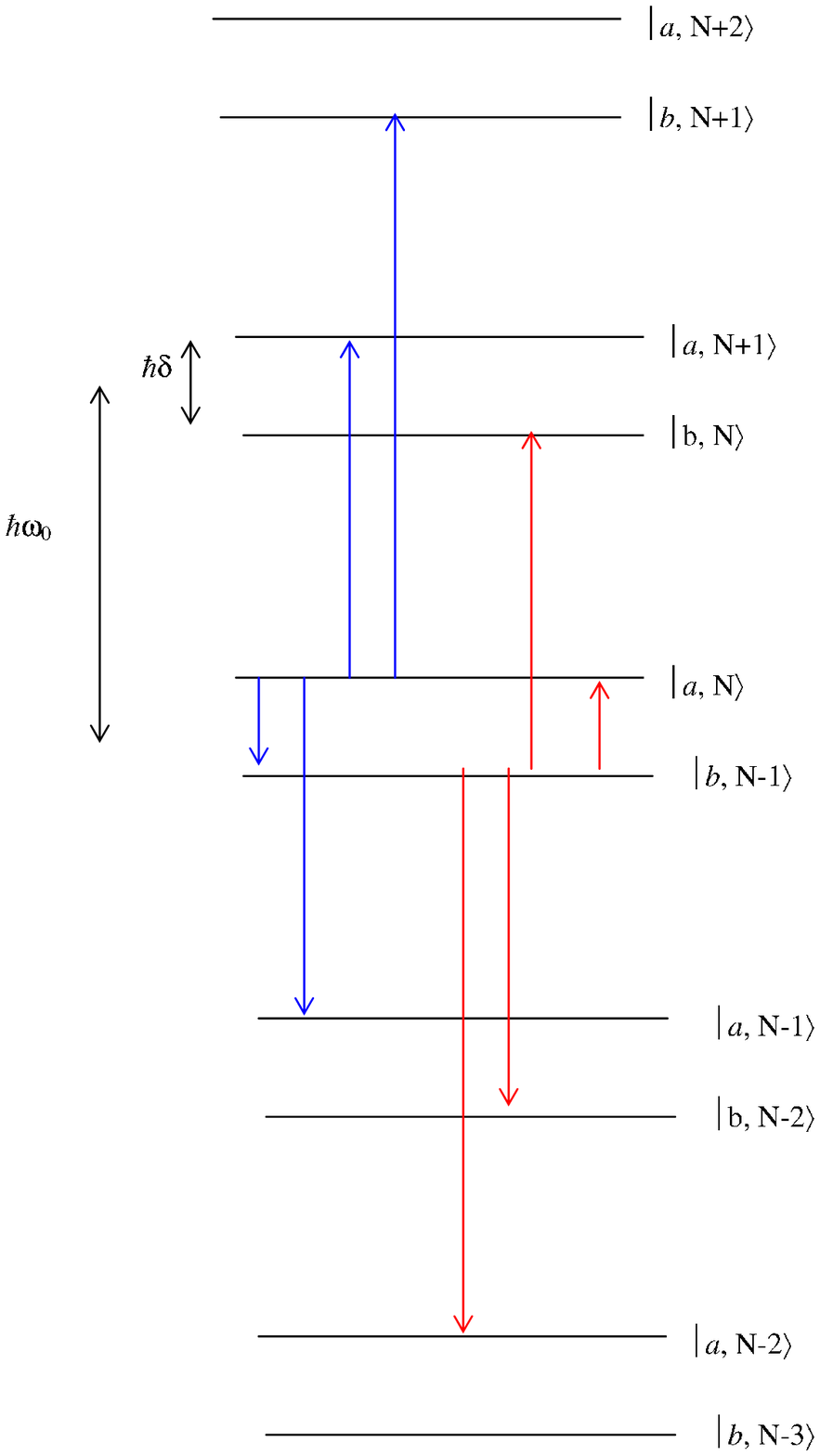}\\
  \caption{(Color online). Levels of the noninteracting TLS and associated photon field system.
  The level spacing in a given manifold is $\hbar\delta$. The spacing between neighboring
  manifolds is $\hbar\omega_0$. The red (blue) arrows show the  transitions  induced by $H_{int}$ from
  the state $|b,N-1\rangle$ ($|a,N\rangle$). }\label{Fig1}
\end{figure}

The Hamiltonian of the TLS in  (\ref{Ham1}) is written in the localized
state basis, i.e. in the basis of states localized in each well.
In terms of the eigenstates basis, which we denote by upper-case
subscripts for the Pauli matrices $\sigma_X$, $\sigma_Y$, and $\sigma_Z$,
Hamiltonian (\ref{Ham1}) reads as:
\begin{equation}\label{Ham3}
H = \frac{\Delta_\varepsilon }{2}\sigma _Z
+\hbar\omega_0(a^+a+1/2)+H_{int}
\end{equation}
where $\Delta_\varepsilon=\sqrt{\Delta^2+\varepsilon^2}$ is the
gap between the two energy states and
\begin{equation}\label{Ham4}
 H_{int} = \frac{1}{2}\left(\frac{\Delta}{\Delta_\varepsilon} \sigma _X- \frac{\varepsilon}{\Delta_\varepsilon} \sigma _Z\right)
 F(a^++a)
 \end{equation}
First, we consider a noninteracting system which involves the TLS and a laser field. The system is
described by the Hamiltonian:
\begin{equation}\label{Ham5}
 H_0 = \frac{\Delta_\varepsilon }{2}\sigma
_Z +\hbar\omega_0(a^+a+1/2)
\end{equation}
We denote  the ground state and
excited state wave functions of the TLS as $|a\rangle$ and $|b\rangle$, respectively, where they have
properties: $\sigma_Z|a\rangle=-|a\rangle$,
$\sigma_Z|b\rangle=|b\rangle$, $\sigma_X|a\rangle=|b\rangle$,
$\sigma_X|b\rangle=|a\rangle$. The eigenfunctions of the photon
field are $|N\rangle$: $a^+|N\rangle=\sqrt{N+1}|N+1\rangle$, and
$a|N\rangle=\sqrt{N}|N-1\rangle$. We denote the eigenfunctions of
the noninteracting TLS and associated photon system as a tensor product
$|a,N\rangle\equiv|a\rangle\otimes|N\rangle$, and
$|b,N\rangle\equiv|b\rangle\otimes|N\rangle$. Up to a constant
term the energies of these states are:
\begin{equation}\label{Ea}
    H_0|a,N\rangle=\left(-\frac{\Delta_\varepsilon}{2}+\hbar\omega_0N\right)|a,N\rangle
\end{equation}

\begin{equation}\label{Eb}
    H_0|b,N\rangle=\left(\frac{\Delta_\varepsilon}{2}+\hbar\omega_0N\right)|b,N\rangle
\end{equation}
For a photon frequency $\omega_0$ close to the TLS frequency
$\Delta_\varepsilon/\hbar$ and a small detuning
$\delta=\omega_0-\Delta_\varepsilon/\hbar\ll \omega_0,
\Delta_\varepsilon/\hbar$, where for definitiveness we assume
$\delta>0$, it is seen from (\ref{Ea}) and (\ref{Eb}) that
the energies of the states $|a,N+1\rangle$ and $|b,N\rangle$ are
close to each other: $E_{a,N+1}-E_{b,N}=\hbar\delta$. The same is
true for the pairs of states $|a,N\rangle$ and $|b,N-1\rangle$,
$|a,N+2\rangle$ and $|b,N+1\rangle$, and so on. Therefore, the
energy levels of the system under concideration is a
ladder of pairs of manifolds which are specified by the photon
number $N$ (see Fig. \ref{Fig1}).
Every manifold is parameterized by a pair of states with a small
spacing between them, $\hbar\delta$, and the distance between
neighboring manifolds is equal to the photon energy, $\hbar\omega_0$.

This ladder of manifolds is quite similar to the one for
atom-field interactions \cite{Coen}. However, a principal
difference is the structure of the interaction Hamiltonian
(\ref{Ham4}). In quantum optics there is no "longitudinal"
interaction between an atomic spin and photon field which is
proportional to $\sigma_Z$. It is the presence of this bias
interaction in a dissipative TLS that leads to some effects which
are not observed in quantum optics.

It should be considered now how these levels are modified due to
the interaction with (\ref{Ham4}). For a pair of these closed spacing
levels within a given manifold, $|a,N\rangle$, $|b,N-1\rangle$
interaction (\ref{Ham4}) causes a transition between them with the
amplitude
\begin{equation}\label{Ta}
    \langle a,N|H_{int}|b,N-1\rangle=\frac{\Delta
    F}{2\Delta_\varepsilon}\sqrt{N}
\end{equation}
Therefore, within 2D Hilbert space the wave functions
$|a,N\rangle$ and $|b,N-1\rangle$ are mixed to give new wave
functions for the dressed states $|1,N\rangle$, and $|2,N\rangle$:
\begin{equation}\label{WF+}
|1,N\rangle=\sin\theta|a,N\rangle+\cos\theta|b,N-1\rangle
\end{equation}

\begin{equation}\label{WF-}
|2,N\rangle=-\cos\theta|a,N\rangle+\sin\theta|b,N-1\rangle
\end{equation}
where we define the state with higher energy as $|1,N\rangle$.

The form of Eqs. (\ref{WF+}) and (\ref{WF-}) ensures the normalization
and orthogonality of the wave functions $|1,N\rangle$ and
$|2,N\rangle$, which are eigenfunctions of Hamiltonian
(\ref{Ham3}). Accordingly, the uncoupled states $|a,N\rangle$,
$|b,N-1\rangle$ can be expressed in terms of the dressed states
$|1,N\rangle$, $|2,N\rangle$:
\begin{equation}\label{WFa}
|a,N\rangle=\sin\theta|1,N\rangle-\cos\theta|2,N\rangle
\end{equation}

\begin{equation}\label{WFb}
|b,N-1\rangle=\cos\theta|1,N\rangle+\sin\theta|2,N\rangle
\end{equation}
By using standard quantum mechanical techniques we can find the
eigenenergies and the angle $\theta$:
\begin{equation}\label{Eig}
    E_{\pm}=\frac{1}{2}\left(E_{|a,N\rangle}+E_{|b,N-1\rangle}\right)\pm\frac{1}{2}\hbar\Omega_R
\end{equation}
where the upper (lower) sign corresponds to $|1,N\rangle$
($|2,N\rangle$). The quantity $\Omega_R$ in (\ref{Eig}) is the
Rabi frequency \cite{com1}
\begin{equation}\label{RF}
   \Omega_R=\sqrt{\delta^2+\Omega_1^2}
\end{equation}
where $\Omega_1=\Delta F/\hbar\Delta_\varepsilon$, and
 $\sqrt{N}$ is incorporated in the high frequency amplitude $F$.

For the angle $\theta$ we obtain $\tan2\theta=-\Omega_1/\delta$,
where $0<2\theta<\pi$, so that $\cos2\theta=-\delta/\Omega_R$,
$\cos\theta=\frac{1}{\sqrt{2}}\left(1-\frac{\delta}{\Omega_R}\right)^{1/2}$, and
$\sin\theta=\frac{1}{\sqrt{2}}\left(1+\frac{\delta}{\Omega_R}\right)^{1/2}$.

\begin{figure}
  \includegraphics[width=8 cm]{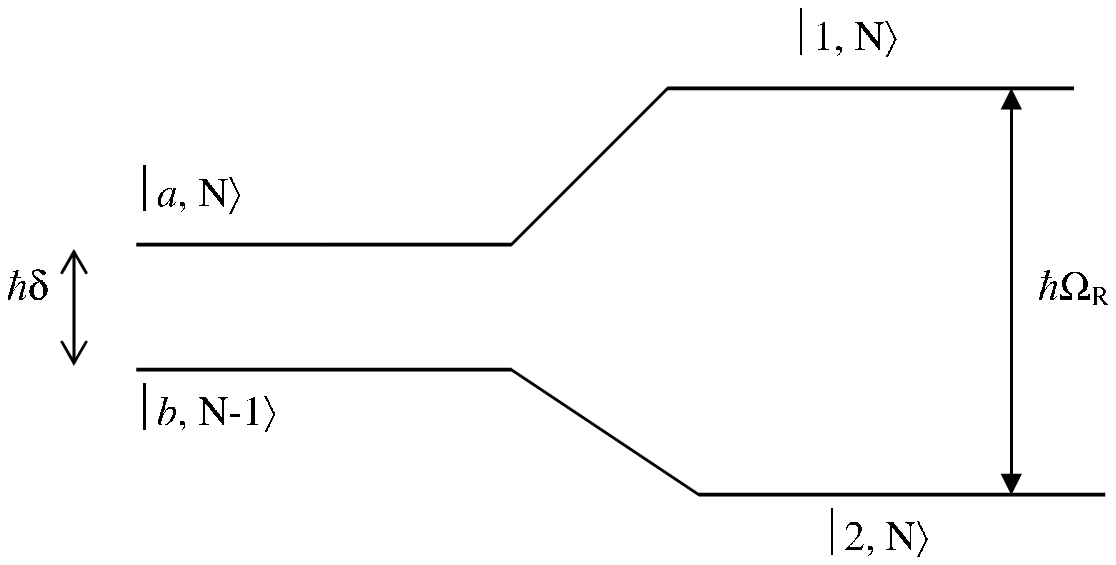}\\
  \caption{The level structure of a dressed manifold.
  The uncoupled states $|a,N\rangle$ and
  $|b,N-1\rangle$ are transformed by the interaction between the
  TLS and photon field into the two dressed states $|1,N\rangle$ and $|2,N\rangle$.}\label{Fig2}
\end{figure}

When the interaction is switched off ($F\rightarrow 0$) then, as
would be expected, the state $|1,N\rangle$ tends to
$|a,N\rangle$, and the state $|2,N\rangle$ tends to
$|b,N-1\rangle$. Therefore, by taking into account the interaction between the TLS and the
photon field the level structure of a given manifold looks like
that shown in Fig.~\ref{Fig2}. The interaction results in an
increase of the energy gap between the states $|a,N\rangle$ and
$|b,N-1\rangle$. Consequently, it is said that these states are dressed by the
interaction. From now on we will call these two nearby dressed
states Rabi levels.

Up to now the picture has been quite similar to that known from
atom-photon interactions\cite{Coen}. However, a drastic difference
appears if we consider the excitation of the dressed levels
$|1,N\rangle$ and $|2,N\rangle$ by a signal the frequency of which
is comparable with Rabi frequency $\Omega_R$. Such a low frequency
signal cannot change the number $N$ of high frequency photons,
therefore, in quantum optics the transition between these two
states is not allowed since, in the language of quantum optics, the atom dipole
operator is transversal: it is proportional to $\sigma_X$.
Therefore, the atom dipole operator connects only the levels, say
$|a,N\rangle$ and $|b,N\rangle$, which belong to the different
manifolds.

In the TLS the transitions between nearby states $|1,N\rangle$ and
$|2,N\rangle$ are caused by the "longitudinal" term in the interaction
Hamiltonian, which is proportional to $\sigma_Z$. Let us assume
that system (\ref{Ham1}) additionally interacts with the low
frequency signal $G\cos\omega t$, so that the term $-\sigma_zG\cos\omega t$
is added to Hamiltonian (\ref{Ham1}), where the frequency
$\omega$ is of the same order of magnitude as the Rabi frequency $\Omega_R$. In
the eigenbasis this low frequency Hamiltonian is transformed as
follows:
\begin{equation}\label{HamLF}
    H_{int}^{LF}=\left(\frac{\Delta}{\Delta_\varepsilon} \sigma _X- \frac{\varepsilon}{\Delta_\varepsilon} \sigma _Z\right)
 G\cos\omega t
\end{equation}
The transitions between the Rabi levels $|1,N\rangle$ and
$|2,N\rangle$ are caused by the second term in the brackets of
(\ref{HamLF}):
\begin{equation}\label{Trans}
    \langle1,N|H_{int}^{LF}|2,N\rangle=-\frac{\varepsilon G\Omega_1}{\Delta_{\varepsilon }\Omega_R}
\end{equation}
In what follows we show that these transitions result in the
undamped low frequency oscillations of the populations of initial
levels $|a\rangle$ and $|b\rangle$ which can be detected by using appropriate electronic circuitry.
\section{Rate equations for the density matrix in uncoupled basis}
The  elements of the density matrix $\sigma$ in the basis of the uncoupled
levels  $|a,N\rangle$ and $|b,N\rangle$ are defined as follows:
\begin{eqnarray}
   \sigma_{aa}^{NN'}&=&\langle a,N|\sigma|a,N'\rangle\nonumber\\
  \sigma_{bb}^{NN'}&=&\langle b,N|\sigma|b,N'\rangle\\
\sigma_{ab}^{NN'}&=&\langle a,N|\sigma|b,N'\rangle\nonumber
\end{eqnarray}
Since the spontaneous transitions between the levels $|a\rangle$
and $|b\rangle$, which lead to their decay do not change the photon
numbers the rate equations for the density matrix $\sigma$ can be
written in the following form:
\begin{equation}\label{bb}
    \frac{d\sigma_{bb}^{NN'}}{dt}=-\frac{i}{\hbar}\left(E_{bN}-E_{bN'}\right)\sigma_{ab}^{NN'}
    -\sigma_{bb}^{NN'}\Gamma_{\downarrow}+\sigma_{aa}^{NN'}\Gamma_{\uparrow}
\end{equation}
\begin{equation}\label{aa}
    \frac{d\sigma_{aa}^{NN'}}{dt}=-\frac{i}{\hbar}\left(E_{aN}-E_{aN'}\right)\sigma_{ab}^{NN'}
    -\sigma_{aa}^{NN'}\Gamma_{\uparrow}+\sigma_{bb}^{NN'}\Gamma_{\downarrow}
\end{equation}

\begin{equation}\label{ab}
    \frac{d\sigma_{ab}^{NN'}}{dt}=-\frac{i}{\hbar}\left(E_{aN}-E_{bN'}\right)\sigma_{ab}^{NN'}-\sigma_{ab}^{NN'}\Gamma_{\varphi}
\end{equation}
where $\Gamma_{\downarrow}$ is the transition rate from the state
$|b,N\rangle$ to state $|a,N\rangle$ (relaxation rate),
$\Gamma_{\uparrow}$ is the transition rate from the state
$|a,N\rangle$ to state $|b,N\rangle$ (excitation rate), and the
quantity $\Gamma_{\varphi}$ is the rate of decoherence. For
equilibrium conditions the relaxation and excitation rates are
related by the detailed balance law:
\begin{equation}\label{balance}
\Gamma_{\uparrow}=\Gamma_{\downarrow}\exp\left(-\frac{\Delta_{\varepsilon}}{k_BT}\right)
\end{equation}
From (\ref{balance})we obtain
\begin{equation}\label{balance1}
   \Gamma_-T_1=-\tanh\left(\frac{\Delta_{\varepsilon}}{2k_BT}\right)
\end{equation}
where we define $\Gamma_-=\Gamma_{\uparrow}-\Gamma_{\downarrow}$
and the longitudinal relaxation time
$T^{-1}_1=\Gamma_{\uparrow}+\Gamma_{\downarrow}$.

The equations (\ref{bb}), (\ref{aa}), and (\ref{ab}) can be written in
the operator form:
\begin{equation}\label{dm}
    \frac{d\sigma}{dt}=-\frac{i}{\hbar}[H_0,\sigma]+\widehat{L}
\end{equation}
where the Hamiltonian of the uncoupled system $H_0$ is given in
(\ref{Ham5}). The operator $\widehat{L}$ is defined by its matrix
elements which follows from (\ref{bb}), (\ref{aa}) and (\ref{ab}):
\begin{multline}\label{L1}
\widehat{L}=-\Gamma_{\downarrow}\sum_{N_1,N_2}|bN_1\rangle\langle
bN_1|\sigma|bN_2\rangle\langle bN_2|\\+
\Gamma_{\uparrow}\sum_{N_1,N_2}|bN_1\rangle\langle
aN_1|\sigma|aN_2\rangle\langle
bN_2|\\-\Gamma_{\uparrow}\sum_{N_1,N_2}|aN_1\rangle\langle
aN_1|\sigma|aN_2\rangle\langle
aN_2|\\+\Gamma_{\downarrow}\sum_{N_1,N_2}|aN_1\rangle\langle
bN_1|\sigma|bN_2\rangle\langle
aN_2|\\-\Gamma_{\varphi}\sum_{N_1,N_2}|aN_1\rangle\langle
aN_1|\sigma|bN_2\rangle\langle
bN_2|\\-\Gamma_{\varphi}\sum_{N_1,N_2}|bN_1\rangle\langle
bN_1|\sigma|aN_2\rangle\langle aN_2|
\end{multline}
Below we assume that the interaction between the TLS and laser field does not influence the $\Gamma$'s rates and the corresponding rate equations. As was shown in \cite{Bloch,Red}, this assumption is valid for relative weak driving, sufficient short correlation time of the heat bath $\tau_c$, and in large temperature limit: $F<<\Delta_{\varepsilon}, \hbar/\tau_c, k_BT$\cite{com}.

\section{Bloch type equations for the reduced density matrix}

Equation (\ref{dm}) can be generalized to include the
interaction between the TLS and laser field:
\begin{equation}\label{dm1}
    \frac{d\sigma}{dt}=-\frac{i}{\hbar}[H,\sigma]+\widehat{L}
\end{equation}
where the Hamiltonian $H$ is given in (\ref{Ham3}) with $H_{int}$ in
(\ref{Ham4}).

Following on, we define the reduced density matrix for the two level coupled
system by tracing over the photon number $N$:
\begin{eqnarray}\label{dm2a}
\rho_{11}=\sum_N\langle 1,N|\sigma|1,N\rangle\nonumber\\
\rho_{22}=\sum_N\langle 2,N|\sigma|2,N\rangle\nonumber\\
\rho_{12}=\sum_N\langle 1,N|\sigma|2,N\rangle\nonumber\\
\rho_{21}=\sum_N\langle 2,N|\sigma|1,N\rangle
\end{eqnarray}

\begin{eqnarray}\label{dm2b}
\kappa_{11}=\sum_N\langle 1,N|\sigma|1,N-1\rangle\nonumber\\
\kappa_{22}=\sum_N\langle 2,N|\sigma|2,N-1\rangle\nonumber\\
\kappa_{12}=\sum_N\langle 1,N|\sigma|2,N-1\rangle\nonumber\\
\kappa_{21}=\sum_N\langle 2,N|\sigma|1,N-1\rangle
\end{eqnarray}
It is convenient to write the rate equations in terms of the new
variables: $\rho=\rho_{11}-\rho_{22}$, $\rho_+=\rho_{12}+\rho_{21}$,
$\rho_-=\rho_{12}-\rho_{21}$, $\kappa^+=\kappa_{11}+\kappa_{22}$,
$\kappa=\kappa_{11}-\kappa_{22}$,
$\kappa_+=\kappa_{12}+\kappa_{21}$,
$\kappa_-=\kappa_{12}-\kappa_{21}$.
It is not difficult to show that the total
population  $\rho^+=\rho_{11}+\rho_{22}$ is constant: $\frac{d\rho^+}{dt}=0$ with
the normalization condition $\rho^+=1$.

By taking into account the matrix elements of $\widehat L$ in the
dressed state basis (see Appendix A), we obtain the rate
equations for the elements of the reduced density matrix:

\begin{equation}\label{ro}
    \frac{d\rho}{dt}=-A_1\rho+B\rho_++\Gamma_-\cos2\theta
\end{equation}

\begin{equation}\label{ro+}
    \frac{d\rho_+}{dt}=-i\Omega_R\rho_-+B\rho-A_2\rho_++\Gamma_-\sin2\theta
\end{equation}

\begin{equation}\label{ro-}
     \frac{d\rho_-}{dt}=-i\Omega_R\rho_+-\Gamma_{\varphi}\rho_-
\end{equation}
\begin{equation}\label{k}
    \frac{d\kappa^+}{dt}=-i\omega_0\kappa^+
\end{equation}

\begin{equation}\label{kap}
    \frac{d\kappa}{dt}=-i\omega_0\kappa-A_1\kappa+B\kappa_++\kappa^+\Gamma_-\cos2\theta
\end{equation}

\begin{equation}\label{kap+}
    \frac{d\kappa_+}{dt}=-i\omega_0\kappa_+-i\Omega_R\kappa_-+B\kappa-A_2\kappa_++\kappa^+\Gamma_-\sin2\theta
\end{equation}

\begin{equation}\label{kap-}
     \frac{d\kappa_-}{dt}=-i\omega_0\kappa_--i\Omega_R\kappa_+-\Gamma_{\varphi}\kappa_-
\end{equation}

where
\begin{equation}\label{A1}
    A_1=\left[\frac{1}{T_1}\cos^22\theta+\Gamma_{\varphi}\sin^22\theta\right]
\end{equation}
\begin{equation}\label{A2}
    A_2=\left[\frac{1}{T_1}\sin^22\theta+\Gamma_{\varphi}\cos^22\theta\right]
\end{equation}
\begin{equation}\label{B}
    B=\left[\Gamma_{\varphi}-\frac{1}{T_1}\right]\sin2\theta\cos2\theta
\end{equation}

The $\rho$'s elements of the density matrix describe the
transitions between the Rabi levels, which are usually accounted for
by a so called rotating wave approximation (RWA). The $\kappa$'s
elements describe the transitions between neighboring manifolds
spaced apart by the energy $\hbar\omega_0$. In the above equations it can be seen that
these two types of transitions are completely
independent: the equations for the $\rho$ matrix are uncoupled from
those for the $\kappa$ matrix.

If the damping is absent (all $\Gamma$'s in (\ref{ro}),
(\ref{ro+}), (\ref{ro-}) are equal to zero) the quantity $\rho$ is
constant, and $\rho_+$ and $\rho_-$ oscillate with the Rabi frequency
$\Omega_R$. However, in the presence of damping these oscillations
rapidly decay to their steady state values.
Here it is instructive to consider two limiting cases.
\subsection{Bloch equations in the absence of a high frequency excitation}
Let the power of high frequency photon field tends to zero. In
this case $\Omega_1\rightarrow 0$, $\omega_0\rightarrow 0$, and
therefore, $\sin2\theta\rightarrow 0$, $\cos2\theta\rightarrow
-1$, and $\Omega_R\rightarrow \Delta_{\varepsilon}/\hbar$.  From
Eqs. (\ref{ro}), (\ref{ro+}), (\ref{ro-}) we get:
\begin{equation}\label{ro0*}
    \frac{d\rho}{dt}=-\rho\frac{1}{T_1}-\Gamma_-
\end{equation}

\begin{equation}\label{ro+*}
    \frac{d\rho_+}{dt}=-i\frac{\Delta_{\varepsilon}}{\hbar}\rho_--\Gamma_{\varphi}\rho_+
\end{equation}

\begin{equation}\label{ro-*}
    \frac{d\rho_-}{dt}=-i\frac{\Delta_{\varepsilon}}{\hbar}\rho_+-\Gamma_{\varphi}\rho_-
\end{equation}
This result is obvious: the ladder of manifolds is reduced to just
two levels $|a\rangle$ and $|b\rangle$. The diagonal part of
the density matrix relaxes with a rate of $1/T_1$ to its steady state value
$\rho^{(0)}=-T_1\Gamma_-$, while the offdiagonal elements $\rho_\pm$
exhibit the damped oscillations with a rate $\Gamma_{\varphi}$
and frequency $\Delta_{\varepsilon}/\hbar$.

\subsection{Bloch equations for zero high frequency detuning {$\delta$}.}
Consider the case when the high frequency detuning  is zero
($\delta=0$). In this limit $\sin2\theta\rightarrow 1$,
$\cos2\theta\rightarrow 0$ and we get from the Eqs. (\ref{ro}),
(\ref{ro+}), (\ref{ro-}):
\begin{equation}\label{ro0}
    \frac{d\rho}{dt}=-\Gamma_{\varphi}\rho
\end{equation}

\begin{equation}\label{ro+0}
    \frac{d\rho_+}{dt}=-i\Omega_1\rho_--\frac{1}{T_1}\rho_++\Gamma_-
\end{equation}

\begin{equation}\label{ro-0}
    \frac{d\rho_-}{dt}=-i\Omega_1\rho_+-\Gamma_{\varphi}\rho_-
\end{equation}
It is seen that the high frequency excitation drastically changes
the behavior of the density matrix. Here the population $\rho$ decays
with the decoherence rate $\Gamma_{\varphi}$. This is due to the
fact that the population of the level, say, $|1,N\rangle$ can only
be changed as a result of spontaneous transitions to the levels
$|1,N-1\rangle$ and  $|2,N-1\rangle$ of the neighboring manifold. The
offdiagonal quantities $\rho_+$ and $\rho_-$ undergo the damped
oscillations with a rate
$\frac{1}{2}\left(\frac{1}{T_1}+\Gamma_{\varphi}\right)$ provided
that
$\Omega_1>\frac{1}{2}\left(\frac{1}{T_1}-\Gamma_{\varphi}\right)$.

\subsection{Steady state solution for the density matrix}
The steady state solution
($\frac{d\rho}{dt}=\frac{d\rho_-}{dt}=\frac{d\rho_+}{dt}=0$) for
Eqs. (\ref{ro}), (\ref{ro+}), (\ref{ro-}) is as follows:
\begin{equation}\label{st0}
    \rho^{(0)}=\frac{\left(\Gamma_{\varphi}^2+\Omega_R^2\right)}{\frac{\Gamma_{\varphi}^2}{T_1}+A_1\Omega_R^2}\Gamma_-\cos2\theta
\end{equation}

\begin{equation}\label{st+}
    \rho_+^{(0)}=\frac{\Gamma_{\varphi}^2}{\frac{\Gamma_{\varphi}^2}{T_1}+A_1\Omega_R^2}\Gamma_-\sin2\theta
\end{equation}

\begin{equation}\label{st-}
    \rho_-^{(0)}=-i\frac{\Omega_R}{\Gamma_{\varphi}}\rho_+^{(0)}
\end{equation}
It is interesting to note that under high frequency irradiation
the population of the Rabi levels becomes inverted. In
(\ref{st0}), the quantity $\rho^{(0)}$, which is the difference of
the populations between higher and lower Rabi levels, becomes
positive, since for $\delta>0$ we have
$\cos2\theta=-\delta/\Omega_R<0$, and always $\Gamma_-<0$.

For the case when the high frequency detuning $\delta$ is small
compared to the Rabi frequency, at zero detuning
($\delta\ll\Omega_1$), we have $\sin2\theta\rightarrow 1$ and
$\cos2\theta\rightarrow -\delta/\Omega_1$, and we get from Eqs.
(\ref{st0}), (\ref{st+}), and (\ref{st-}):
\begin{equation}\label{std0}
    \rho^{(0)}=-\frac{\delta}{\Gamma_{\varphi}\Omega_1}\frac{\Gamma_-\left(\Gamma^2_{\varphi}+\Omega_1^2\right)}
    {\Omega_1^2+\frac{\Gamma_{\varphi}}{T_1}}
\end{equation}

\begin{equation}\label{std+0}
    \rho_+^{(0)}=\frac{\Gamma_-\Gamma_{\varphi}}{\Omega_1^2+\frac{\Gamma_{\varphi}}{T_1}}
\end{equation}

\begin{equation}\label{std-0}
    \rho_-^{(0)}=-i\frac{\Gamma_-\Omega_1}{\Omega_1^2+\frac{\Gamma_{\varphi}}{T_1}}
\end{equation}
As seen from (\ref{std0}) $\rho^{(0)}\rightarrow 0$ as
$\delta$ tends to zero. This causes the equalization of the
population of the two levels ($\rho_{11}=\rho_{22}=\frac{1}{2}$) when
the high frequency is in exact resonance with the energy gap of the TLS.

The steady state solutions of Eqs. (\ref{k}), (\ref{kap}),
(\ref{kap+}), and (\ref{kap-})
 are equal to zero:
$\kappa^{+(0)}=0$, $\kappa^{(0)}=0$, $\kappa_+^{(0)}=0$,
$\kappa_-^{(0)}=0$. This implies  $\kappa_{11}^{(0)}=0$,
$\kappa_{22}^{(0)}=0$, $\kappa_{12}^{(0)}=0$, and
$\kappa_{21}^{(0)}=0$.
\section{Excitation of Rabi levels by a low frequency signal}
In this section we find the response of the TLS coupled to a photon field to an
external signal the frequency of which is of the order of the Rabi
frequency $\Omega_R$. The operator equation for the density matrix
$\sigma$ is similar to (\ref{dm1}):
\begin{equation}\label{dmint}
    \frac{d\sigma}{dt}=-\frac{i}{\hbar}[H+H_{int}^{LF},\sigma]+\widehat{L}
\end{equation}
where the Hamiltonian $H_{int}^{LF}$ is given in (\ref{HamLF}).

Since the low frequency signal cannot change the photon number
$N$, the transitions between Rabi levels $|1,N\rangle$ and
$|2,N\rangle$ can be induced only by the second term in the low
frequency Hamiltonian (\ref{HamLF}). The equations for the $\rho$'s
and the $\kappa$'s are obtained in the same way as used for Eqs. (\ref{ro})-
(\ref{kap-}). The only difference is the appearance of low
frequency terms in the right hand side of these equations. Therefore,
by taking into account the low frequency excitation, we get  the
following Bloch like equations for the reduced density matrix:
\begin{multline}\label{roLF}
    \frac{d\rho}{dt}=-A_1\rho+B\rho_+-
    \rho_-\left(ig_1\sin2\theta\cos\omega t\right)+\\-ig_2\cos\omega t
    \cos^2\theta\left(\kappa_{12}-\kappa_{12}^+\right)+\Gamma_-\cos2\theta
\end{multline}

\begin{equation}\label{roLF1}
    \frac{d\rho^+}{dt}=
    -ig_2\cos\omega t
    \sin^2\theta\left(\kappa_{21}-\kappa_{21}^+\right)
\end{equation}

\begin{multline}\label{roLF2}
    \frac{d\rho_+}{dt}=-i\Omega_R\rho_-+B\rho-A_2\rho_++\rho_-\left(ig_1\cos2\theta\cos\omega t\right)\\
    +ig_2\cos\omega
    t[\left(\kappa_{22}^+-\kappa_{22}\right)+\left(\kappa_{11}^+-\kappa_{11}\right)\cos2\theta+\\
    \left(\kappa_{21}+\kappa_{12}^+-\kappa_{21}^+-\kappa_{12}\right)\sin2\theta]+\Gamma_-\sin2\theta
\end{multline}

\begin{multline}\label{roLF3}
     \frac{d\rho_-}{dt}=-i\Omega_R\rho_+-\Gamma_{\varphi}\rho_-+
     ig_1\left(\rho_+\cos2\theta-\rho\sin2\theta\right)\cos\omega
     t\\-ig_2\cos\omega
     t[\sin2\theta\left(\kappa_{12}+\kappa^+_{12}+\kappa_{21}+\kappa^+_{21}\right)\\-
     \left(\kappa_{11}+\kappa^+_{11}\right)-\left(\kappa_{22}+\kappa^+_{22}\right)\cos2\theta]
\end{multline}
where $g_1=2\varepsilon G/\hbar\Delta_{\varepsilon}$, and $g_2=2\Delta
G/\hbar\Delta_{\varepsilon}$.

\begin{multline}\label{kappa}
    \frac{d\kappa}{dt}=-i\omega_0\kappa-A_1\kappa+B\kappa_++\kappa^+\Gamma_-\cos2\theta\\+
    ig_2\cos\omega
 t\left[\rho_+-\rho_-\cos2\theta\right]
-ig_1\kappa_-\cos\omega t\sin2\theta
\end{multline}

\begin{multline}\label{kappa+}
    \frac{d\kappa_+}{dt}=-i\omega_0\kappa_+-i\Omega_R\kappa_-+B\kappa-A_2\kappa_++\kappa^+\Gamma_-\sin2\theta\\
-ig_2\cos\omega t\left[\rho_-
\sin2\theta+\rho\right]+ig_1\kappa_-\cos\omega t\cos2\theta
\end{multline}

\begin{multline}\label{kappa-}
     \frac{d\kappa_-}{dt}=-i\omega_0\kappa_--i\Omega_R\kappa_+-\Gamma_{\varphi}\kappa_-\\
     - ig_2\cos\omega
 t\left[\rho_+\sin2\theta+\rho\cos2\theta\right]\\
+ig_1\cos\omega
t\left[\kappa_+\cos2\theta-\kappa\sin2\theta\right]
\end{multline}
The  equation for $\kappa^+$ remains unchanged (Eq. (\ref{k})).
Therefore, the low frequency excitation couples the $\rho$ and
$\kappa$ elements of the reduced density matrix. Due to this coupling the $\kappa$ elements acquire the low frequency part which thus should be included in the standard scheme of RWA.
\section{low frequency linear susceptibilities for the TLS}
It is evident that the above equations exhibit oscillatory
solutions in the presence of damping. For a small amplitude low
frequency signal, $G$ the time dependent solution for Eqs.
(\ref{roLF}), (\ref{roLF1}), (\ref{roLF2}), and (\ref{roLF3}) can be
obtained by using a perturbation method where the small time-dependent
corrections to the steady state values are:
$\rho(t)=\rho^{(0)}+\rho^{(1)}(t)$,
$\rho_+(t)=\rho_+^{(0)}+\rho^{(1)}_+(t)$ , and
$\rho_-(t)=\rho_-^{(0)}+\rho^{(1)}_-(t)$. By doing this we may
neglect all $\kappa$'s on the right hand side of these equations,
since the steady state values for the $\kappa$'s are zero. Therefore,
in this approximation the equations for the time dependent corrections
to the $\rho$'s are decoupled from those to the $\kappa$'s:

\begin{multline}\label{ro1LF}
    \frac{d\rho^{(1)}}{dt}=-A_1\rho^{(1)}+B\rho^{(1)}_+-
    \rho^{(0)}_-\left(ig\sin2\theta\cos\omega t\right)
\end{multline}

\begin{multline}\label{ro+1LF}
    \frac{d\rho^{(1)}_+}{dt}=-i\Omega_R\rho^{(1)}_-+
    B\rho^{(1)}-A_2\rho^{(1)}_++
\rho^{(0)}_-\left(ig\cos2\theta\cos\omega t\right)
\end{multline}

\begin{multline}\label{ro-1LF}
     \frac{d\rho^{(1)}_-}{dt}=-i\Omega_R\rho^{(1)}_+-\Gamma_{\varphi}\rho^{(1)}_-+\\
     ig\left(\rho^{(0)}_+\cos2\theta-\rho^{(0)}\sin2\theta\right)\cos\omega t
\end{multline}
where $\rho^{(0)}$, $\rho_+^{(0)}$, and $\rho_-^{(0)}$ are the steady
state values given in (\ref{st0}), (\ref{st+}), and (\ref{st-}).

 From these equations it is not difficult to find the linear
susceptibilities of the system
($\chi_{\rho}(\omega)=\rho(\omega)/G(\omega)$, etc.):
\begin{widetext}
\begin{equation}\label{hiro}
    \chi_{\rho}(\omega)=-\frac{2\varepsilon\Omega_R}{D(\omega)\hbar\Gamma_{\varphi}\Delta_{\varepsilon}}
   \rho_+^{(0)}\left[\sin2\theta\left[\left(i\omega+\Gamma_{\varphi}\right)\left(i\omega+\frac{1}{T_1}\right)+\Omega^2_R\right]
   +\frac{\Omega^2_R}{\Gamma_{\varphi}}B\cos2\theta\right]
\end{equation}

\begin{equation}\label{hiro+}
\chi_{\rho_+}(\omega)=\frac{2\varepsilon\Omega_R}{D(\omega)\hbar\Gamma_{\varphi}\Delta_{\varepsilon}}\cos2\theta
  \rho_+^{(0)}\left[\left(i\omega+\Gamma_{\varphi}\right)\left(i\omega+\frac{1}{T_1}\right)
  -\left(i\omega+A_1\right)\frac{\Omega_R^2}{\Gamma_{\varphi}}\right]
\end{equation}

\begin{equation}\label{hiro-}
\chi_{\rho_-}(\omega)=-i\frac{2\varepsilon\Omega_R^2}{D(\omega)\hbar\Gamma_{\varphi}^2\Delta_{\varepsilon}}\rho_+^{(0)}\cos2\theta
\left(i\omega+\frac{1}{T_1}\right)\left(i\omega+2\Gamma_{\varphi}\right)
\end{equation}
where
\begin{equation}\label{Domega}
    D(\omega)=\left(i\omega+\Gamma_{\varphi}\right)^2
    \left(i\omega+\frac{1}{T_1}\right)+\left(i\omega+A_1\right)\Omega_R^2
\end{equation}
\end{widetext}
Eqs. (\ref{hiro}), (\ref{hiro+}), and (\ref{hiro-}) give the
response of the coupled system (TLS and associated photon field) to a low frequency
signal, which excites transitions between the Rabi levels.

From (\ref{hiro}), (\ref{hiro+}), and (\ref{hiro-}) the linear
susceptibilities for the case of small high frequency detuning,
$\sin2\theta\rightarrow 1$,
$\cos2\theta\rightarrow-\delta/\Omega_1$,
$A_1\rightarrow\Gamma_{\varphi}$, $A_2\rightarrow1/T_1$, and
$B\rightarrow-\frac{\delta}{\Omega_1}\left(\Gamma_{\varphi}-\frac{1}{T_1}\right)$,
can also be obtained.
\begin{equation}\label{hi}
    \chi_{\rho}(\omega)=-\frac{2\varepsilon\Omega_1}{\hbar\Delta_{\varepsilon}\Gamma_{\varphi}}
    \frac{\rho_+^{(0)}}{i\omega+\Gamma_{\varphi}}
\end{equation}

\begin{equation}\label{hi+}
    \chi_{\rho_+}(\omega)=-\delta\frac{2\varepsilon}{\hbar\Delta_{\varepsilon}}
    \frac{\rho_+^{(0)}}{\Gamma_{\varphi}d(\omega)}
    \left[i\omega+\frac{1}{T_1}-\frac{\Omega_1^2}{\Gamma_{\varphi}}\right]
\end{equation}

\begin{multline}\label{hi-}
    \chi_{\rho_-}(\omega)=i\delta\frac{2\varepsilon}{\hbar\Delta_{\varepsilon}}
    \frac{\Omega_1\rho_+^{(0)}}{\Gamma_{\varphi}^2
    d(\omega)}
\frac{\left(i\omega+\frac{1}{T_1}\right)\left(i\omega+2\Gamma_{\varphi}\right)
}{\left(i\omega+\Gamma_{\varphi}\right)}
\end{multline}
where
\begin{equation}\label{dom}
    d(\omega)=\left(i\omega+\frac{1}{T_1}\right)\left(i\omega+\Gamma_{\varphi}\right)+
    \Omega_1^2
\end{equation}
and $\rho_+^{(0)}$ is given by (\ref{std+0}). The resonance nature
of the response is evident from (\ref{dom}).

The equations for the time dependent $\kappa$'s in the first order
in $g$ are coupled to $\rho$'s via their steady state values as
follows:
\begin{equation}\label{k1}
    \frac{d\kappa^{+}}{dt}=-i\omega_0\kappa^{+}
\end{equation}
\begin{multline}\label{kappa1}
    \frac{d\kappa}{dt}=-i\omega_0\kappa-A_1\kappa+B\kappa_++\kappa^+\Gamma_-\cos2\theta\\+
    i\frac{g_2}{2}f_1\cos\omega t
\end{multline}
\begin{multline}\label{kappa+1}
\frac{d\kappa_+}{dt}=-i\omega_0\kappa_+-i\Omega_R\kappa_-+B\kappa-A_2\kappa_++\kappa^+\Gamma_-\sin2\theta\\
-i\frac{g_2}{2}f_2\cos\omega t
\end{multline}
\begin{multline}\label{kappa-1}
     \frac{d\kappa_-}{dt}=-i\omega_0\kappa_--i\Omega_R\kappa_+-\Gamma_{\varphi}\kappa_-
     - i\frac{g_2}{2}f_3\cos\omega t
\end{multline}
where
\begin{equation}\label{f1}
    f_1=\left(\rho_+^{(0)}-\rho_-^{(0)}\cos2\theta\right)
\end{equation}

\begin{equation}\label{f2}
  f_2=\left(\rho_-^{(0)}\sin2\theta+\rho^{(0)}\right)
\end{equation}

\begin{equation}\label{f3}
    f_3=\left(\rho_+^{(0)}\sin2\theta+\rho^{(0)}\cos2\theta\right)
\end{equation}
\begin{widetext}
 The low frequency linear susceptibilities for the
$\kappa$'s ($\chi_{\kappa}(\omega)=\kappa(\omega)/G(\omega)$,
etc.) are as follows: $\chi_{\kappa^+}(\omega)=0$,

\begin{equation}\label{chikappa-}
\chi _{\kappa _ -  } (\omega ) = \frac{\Delta }{{\hbar \Delta
_\varepsilon  D_0(\omega)}}\left( { - if_3 \left[ {i(\omega +
\omega _0 ) + A_1 } \right]\left[ {i(\omega  + \omega _0 ) + A_2 }
\right] + f_1 B\Omega _R  + if_3 B^2  - f_2 \Omega _R \left[
{i(\omega  + \omega _0 ) + A_1 } \right]} \right)
\end{equation}

\begin{equation}\label{chikappa}
\chi _\kappa  (\omega ) = \frac{\Delta }{{\hbar \Delta
_\varepsilon  D_0(\omega)}}\left( {if_1 \left[ {i(\omega + \omega
_0 ) + A_2 } \right]\left[ {i(\omega  + \omega _0 ) + \Gamma
_\varphi  } \right] - if_2 B\left[ {i(\omega  + \omega _0 ) +
\Gamma _\varphi  } \right] - f_3 B\Omega _R  + if_3 \Omega _R^2 }
\right)
\end{equation}
\begin{equation}\label{chikappa+}
\chi _{\kappa _ +  } (\omega ) = \frac{\Delta }{{\hbar \Delta
_\varepsilon  D_0(\omega)}}\left( { - if_2 \left[ {i(\omega +
\omega _0 ) + A_1 } \right]\left[ {i(\omega  + \omega _0 ) +
\Gamma _\varphi  } \right] + if_1 B\left[ {i(\omega  + \omega _0 )
+ \Gamma _\varphi  } \right]   - f_3 \Omega _R \left[ {i(\omega  +
\omega _0 ) + A_1 } \right]} \right)
\end{equation}
where

\begin{equation}\label{Det}
       D_0(\omega)=\left[i(\omega+\omega_0)+\Gamma_{\varphi}\right]^2
    \left(i(\omega+\omega_0)+\frac{1}{T_1}\right)+\left[i(\omega+\omega_0)+A_1\right]\Omega_R^2
\end{equation}
\end{widetext}
By assuming that the gap frequency $\omega_0$ is much more than
$\omega$ and the rates $\Gamma$'s, we obtain from
(\ref{chikappa-}), (\ref{chikappa}), and (\ref{chikappa+}) the
corresponding low frequency susceptibilities in the second order
of the inverse frequency $\omega_0$:
\begin{equation}\label{chikappa-approx}
\chi _{\kappa _ -  } (\omega ) =  - \frac{\Delta }{{\hbar \omega
_0 \Delta _\varepsilon  }}f_3  + \frac{\Delta }{{\hbar \omega _0^2
\Delta _\varepsilon  }}\left( {-i\Gamma_{\varphi}f_3  + \omega f_3
+ f_2 \Omega _R } \right)
\end{equation}

\begin{equation}\label{chikappaapprox}
\chi _\kappa  (\omega ) = \frac{\Delta }{{\hbar \omega _0 \Delta
_\varepsilon  }}f_1  +i\frac{\Delta }{{\hbar \omega _0^2 \Delta
_\varepsilon  }}\left( {A_1f_1  + f_2 B} \right)
\end{equation}

\begin{multline}\label{chikappa+approx}
\chi _{\kappa _ +  } (\omega ) =  - \frac{\Delta }{{\hbar \omega
_0 \Delta _\varepsilon  }}f_2 \\ + \frac{\Delta }{{\hbar \omega
_0^2 \Delta _\varepsilon  }}\left(-iA_2f_2  + \omega f_2 + f_3
\Omega _R - if_1 B \right)
\end{multline}

\section{The applications}
In possible applications of the method that we propose here the
quantities to be measured are the averages of the Pauli spin operators
$\langle\sigma_X\rangle$, $\langle\sigma_Y\rangle$,
$\langle\sigma_Z\rangle$, and their time derivatives
$\frac{d\langle\sigma_X\rangle}{dt}$,
$\frac{d\langle\sigma_Y\rangle}{dt}$,
$\frac{d\langle\sigma_Z\rangle}{dt}$. By using the definitions of
the density matrix (\ref{dm2a}) and (\ref{dm2b}), and the dressed
states (\ref{WF+}) and (\ref{WF-}) we obtain from the direct
application of Eqs. (\ref{Trsigmaz1}) and (\ref{Trsigmax}) (see
Appendix B):

\begin{equation}\label{sigma-z}
     \langle\sigma_Z\rangle= \rho(t)\cos2\theta+\rho_+(t)\sin2\theta
\end{equation}
\begin{equation}\label{sigma-x}
     \langle\sigma_X\rangle=
     \sin2\theta Re[\kappa(t)]-\cos2\theta Re [\kappa_+(t)]-Re [\kappa_-(t)]
\end{equation}
\begin{equation}\label{sigma-y}
     \langle\sigma_Y\rangle=-\sin2\theta Im[\kappa(t)]
     +\cos2\theta Im[\kappa_+(t)]+Im[\kappa_-(t)]
\end{equation}
For a small amplitude low frequency excitation $G\cos\omega t$
we calculate, below, the time derivatives of (\ref{sigma-z}), (\ref{sigma-x}), and
(\ref{sigma-y}) with the help of equations
(\ref{ro1LF}), (\ref{ro+1LF}), and (\ref{ro-1LF}) for the $\rho$'s and
(\ref{kappa1}), (\ref{kappa+1}), and (\ref{kappa-1}) for the $\kappa$'s:

\begin{multline}\label{dsxdt}
    \frac{d\langle\sigma_X\rangle}{dt}=
  \omega_0\sin2\theta Im [\kappa(t)]-\Gamma_{\varphi}\sin2\theta Re[\kappa(t)]\\
  +\Gamma_{\varphi}\cos2\theta Re[\kappa_+(t)]
  -\left(\omega_0+\Omega_R\cos2\theta\right)Im[\kappa_-(t)]\\
-\left(\omega_0\cos2\theta+\Omega_R\right)Im[\kappa_+(t)]+\Gamma_{\varphi}Re[\kappa_-(t)]
\end{multline}
\begin{multline}\label{dsydt}
    \frac{d\langle\sigma_Y\rangle}{dt}=
  \omega_0\sin2\theta Re[\kappa(t)]+\Gamma_{\varphi}\sin2\theta Im [\kappa(t)]\\
  -\Gamma_{\varphi}\cos2\theta Im [\kappa_+(t)]
  -\left(\omega_0+\Omega_R\cos2\theta\right)Re[\kappa_-(t)]\\
-\left(\omega_0\cos2\theta+\Omega_R\right)Re[\kappa_+(t)]-\Gamma_{\varphi}Im
[\kappa_-(t)] -g_2f_3\cos\omega t
\end{multline}
\begin{equation}\label{dszdt1}
    \frac{d\langle\sigma_Z\rangle}{dt}=-\frac{1}{T_1}\langle\sigma_Z\rangle-i\Omega_R\sin2\theta\rho_-^{(0)}-
    i\Omega_R\sin2\theta\rho_-^{(1)}(t)+\Gamma_-
\end{equation}
where $\langle\sigma_Z\rangle$ in the right hand side of (\ref{dszdt1}) is written using the first approximation only:
\begin{equation}\label{SgZ}
      \langle\sigma_Z\rangle=\rho^{(0)}\cos2\theta+\rho^{(0)}_+\sin2\theta
      +\rho^{(1)}(t)\cos2\theta+\rho^{(1)}_+(t)\sin2\theta
\end{equation}

 The
quantities $\rho^{(1)}(t)$, $\rho^{(1)}_+(t)$, and
$\rho^{(1)}_-(t)$ can be expressed in terms of the real ($\chi'$) and
imaginary ($\chi''$) parts of their corresponding low frequency
susceptibilities: $\rho^{(1)}(t)=G\left(\chi'_{\rho}\cos\omega
t-\chi''_{\rho}\sin\omega t\right)$,
$\rho^{(1)}_+(t)=G\left(\chi'_{\rho_+}\cos\omega
t-\chi''_{\rho_+}\sin\omega t\right)$, and
$\rho^{(1)}_-(t)=iG\left(\chi'_{\rho_-}\sin\omega
t+\chi''_{\rho_-}\cos\omega t\right)$.
\subsection{Flux qubit}
Our method can be directly applied to a persistent current qubit
(flux qubit), which is a superconducting loop interrupted by three
Josephson junctions ~\cite{Mooij,Orlando}. For these qubits the
successful experimental implementation of the low frequency readout
electronics has been demonstrated ~\cite{Green3, Grajcar, Ilich2}.
Though the main subject of the paper is the combined effect on the TLS
of the high and low frequencies, it is worth noting that the pure
low frequency probing of the TLS can also be a valuable tool for
characterization of the two-level systems. In fact, for the flux qubit
this method has been realized theoretically and experimentally ~\cite{Green3, Grajcar}.

\begin{figure}
\includegraphics[width=8cm]{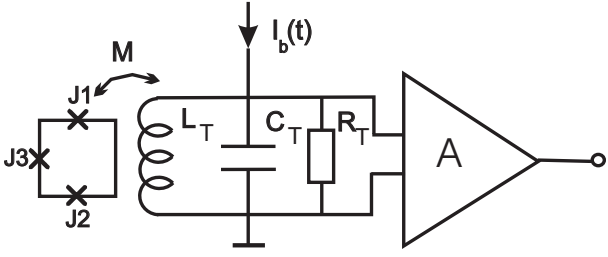}
\caption{\small Flux qubit coupled to a high quality resonant
tank circuit.}\label{Fig3}
\end{figure}

The basis qubit states have opposing persistent currents. The
operator of the persistent current in the qubit loop is
$\widehat{I}_q=I_q\sigma_z$. In eigenstate basis the average
current, $\langle{\widehat{I}_q\rangle}$ is:
\begin{equation}\label{Curr}
\langle \widehat{I}_q\rangle  = \frac{{I_q }}{\Delta_{\varepsilon}
}\left( {\varepsilon \langle {\sigma _Z } \rangle  - \Delta
\langle {\sigma _X } \rangle } \right).
\end{equation}
This current can be detected by a high quality resonant tank
circuit inductively coupled to the qubit loop ~\cite{Ilich2,
Green3, Grajcar}(see Fig. \ref{Fig3}).
For the flux qubit the bias $\varepsilon$ is controlled by an
external dc flux $\Phi_X$: $\varepsilon=E_J(\Phi_X/\Phi_0-1/2)$,
where $E_J=\Phi_0I_q/2\pi$, and $\Phi_0=h/2e$ is the flux quantum.
The qubit is inductively coupled through the mutual inductance
$M=k(L_qL_T)^{1/2}$, where $k$ is the dimensionless coupling
parameter, and $L_q$ is the inductance of the qubit loop, to a
high quality resonant tank circuit with inductance $L_T$,
capacitance $C_T$, and quality factor $Q_T$. The tank circuit is biased by
a low (compared to the gap) frequency current $I_b=I_0\cos\omega
t$. In addition, a high frequency signal $F\cos\omega_0 t$ (not
shown in Fig. \ref{Fig3}) tuned to the gap frequency
$\Delta_{\varepsilon}/\hbar$ is applied to the qubit loop. This
readout circuit has prove to be successful for the investigation
of the quantum properties of the flux qubit ~\cite{Ilich2, Green3,
Grajcar}. The interaction between the qubit and the tank circuit is
described by the term $G(t)=MI_qI_T(t)$, where $I_q$ is the
current flowing in the qubit loop, and $I_T$ is the current in the
tank coil. The voltage across the tank, $V(t)=V_T\cos(\omega t+
\chi)$, is given by ~\cite{Green3}:
\begin{equation}\label{Volt}
    \ddot V + \gamma _T \dot V + \omega _T^2 V = -M\omega _T^2 \frac{{d\langle\widehat{I}_q \rangle}}{{dt}} +
    \omega _T^2 L_T \dot I_b,
\end{equation}
where  $\gamma_T= \omega_T/Q_T$,  and $\omega_T=(L_TC_T)^{-1/2}$
is the tank resonance frequency, which is tuned to the Rabi
frequency $(\omega_T\simeq\Omega_R)$. By using (\ref{dsxdt}) and (\ref{dszdt1})
 we find:
\begin{widetext}
\begin{multline}\label{dotcurr}
\frac{{d\langle\widehat{I}_q
\rangle}}{{dt}}=-\frac{{I_q\varepsilon
}}{\Delta_{\varepsilon}T_1}\left[\rho(t)\cos2\theta+\rho_+(t)\sin2\theta
+i\Omega_RT_1\sin2\theta\rho_-(t)\right]-\frac{{I_q\Delta}}{\Delta_{\varepsilon}}\{\omega_0
\sin2\theta Im[\kappa(t)]-\Gamma_{\varphi}\sin2\theta
Re[\kappa(t)]+\Gamma_{\varphi}Re[\kappa_-(t)]\\
+\Gamma_{\varphi}\cos2\theta Re[\kappa_+(t)]
-\left(\omega_0+\Omega_R\cos2\theta\right)Im [\kappa_-(t)]
-\left(\omega_0\cos2\theta+\Omega_R\right)Im
[\kappa_+(t)]\}\\
\end{multline}

In order to write down the Fourier component of (\ref{dotcurr}) we
use the relations
$\left(Im[k(t)]\right)_\omega=\frac{1}{2i}\left(\chi_{\kappa}(\omega)-\chi^*_{\kappa}(-\omega)\right)G(\omega)$,
$\left(Re[k(t)]\right)_\omega=\frac{1}{2}\left(\chi_{\kappa}(\omega)+\chi^*_{\kappa}(-\omega)\right)G(\omega)$,
etc... Therefore,

\begin{multline}\label{dotcurr1}
\left(\frac{{d\langle\widehat{I}_q
\rangle}}{{dt}}\right)_\omega=-\frac{{I_q\varepsilon G(\omega)
}}{\Delta_{\varepsilon}T_1}\left[\chi_{\rho}(\omega)\cos2\theta+\chi_{\rho_+}(\omega)\sin2\theta
+i\Omega_RT_1\sin2\theta\chi_{\rho_-}(\omega)\right] -\frac{{I_q
\Delta G(\omega)}}{2\Delta_{\varepsilon}}\{-i\omega_0 \sin2\theta
\left[\chi_{\kappa}(\omega)-\chi^*_{\kappa}(-\omega)\right]\\
+\Gamma_{\varphi}\left[\chi_{\kappa_-}(\omega)+\chi^*_{\kappa_-}(-\omega)\right]
-\Gamma_{\varphi}\sin2\theta\left[\chi_{\kappa}(\omega)+\chi^*_{\kappa}(-\omega)\right]
+\Gamma_{\varphi}\cos2\theta\left[\chi_{\kappa_+}(\omega)+\chi^*_{\kappa_+}(-\omega)\right]\\
+i\left(\omega_0+\Omega_R\cos2\theta\right)\left[\chi_{\kappa_-}(\omega)-\chi^*_{\kappa_-}(-\omega)\right]
+i\left(\omega_0\cos2\theta+\Omega_R\right)\left[\chi_{\kappa_+}(\omega)-\chi^*_{\kappa_+}(-\omega)\right]\}\\
\end{multline}

where the expressions for the susceptibilities are given in
(\ref{hiro}), (\ref{hiro+}), (\ref{hiro-}), (\ref{chikappa-}),
(\ref{chikappa}), and (\ref{chikappa+}).

Eq. (\ref{Volt}) in terms of its Fourier components reads:
\begin{equation}\label{Vomega}
  V(\omega )\left( {\omega _T^2  - \omega ^2  + i\omega \gamma _T } \right) =
  -M\omega _T^2 \left(\frac{{d\langle\widehat{I}_q
  \rangle}}{{dt}}\right)_{\omega}
  +i\omega \omega _T^2 L_TI_0
\end{equation}
By taking into account that $G(\omega)=MI_qI_T(\omega)$, where
$I_T(\omega)=-iV(\omega)/\omega L_T$, we obtain for the low
frequency detuning $\xi$, and friction, $\Gamma_T$:

\begin{multline}\label{ksi}
    \xi  = \omega _T^2  - \omega ^2-\frac{{k^2 \omega _T L_q
    I_q^2\varepsilon}}{{\Delta_{\varepsilon}T_1}}\left(\Omega_RT_1\sin2\theta\chi'_{\rho_-}(\omega)
    +\cos2\theta\chi''_{\rho}(\omega)+\sin2\theta\chi''_{\rho_+}(\omega)\right)\\
-\frac{{k^2\omega_TL_qI_q^2}\Delta}{{2\Delta_{\varepsilon}}}
\left(-\omega_0\sin2\theta\left[\chi'_{\kappa}(\omega)-\chi'_{\kappa}(-\omega)\right]
+\Gamma_{\varphi}\left[\chi''_{\kappa_-}(\omega)-\chi''_{\kappa_-}(-\omega)\right]
-\Gamma_{\varphi}\sin2\theta\left[\chi''_{\kappa}(\omega)-\chi''_{\kappa}(-\omega)\right]
\right.\\
+\Gamma_{\varphi}\cos2\theta\left[\chi''_{\kappa_+}(\omega)-\chi''_{\kappa_+}(-\omega)\right]
    \left.+
\left(\omega_0\cos2\theta+\Omega_R\right)
\left[\chi'_{\kappa_+}(\omega)-\chi'_{\kappa_+}(-\omega)\right]
+\left(\omega_0+\Omega_R\cos2\theta\right)
\left[\chi'_{\kappa_-}(\omega)-\chi'_{\kappa_-}(-\omega)\right]\right)
\end{multline}
\begin{multline}\label{GammaT}
    \Gamma _T  = \gamma _T-\frac{{k^2 L_q
    I_q^2\varepsilon}}{\Delta_{\varepsilon}T_1}\left(\Omega_RT_1\sin2\theta\chi''_{\rho_-}(\omega)
     -\cos2\theta\chi'_{\rho}(\omega)-\sin2\theta\chi'_{\rho_+}(\omega)\right)\\
+\frac{{k^2L_qI_q^2}\Delta}{{2\Delta_{\varepsilon}}}
\left(\omega_0\sin2\theta\left[\chi''_{\kappa}(\omega)+\chi''_{\kappa}(-\omega)\right]
+\Gamma_{\varphi}\left[\chi'_{\kappa_-}(\omega)+\chi'_{\kappa_-}(-\omega)\right]
-\Gamma_{\varphi}\sin2\theta\left[\chi'_{\kappa}(\omega)+\chi'_{\kappa}(-\omega)\right]
\right.\\
+\Gamma_{\varphi}\cos2\theta\left[\chi'_{\kappa_+}(\omega)+\chi'_{\kappa_+}(-\omega)\right]
    \left.-
\left(\omega_0\cos2\theta+\Omega_R\right)
\left[\chi''_{\kappa_+}(\omega)+\chi''_{\kappa_+}(-\omega)\right]
-\left(\omega_0+\Omega_R\cos2\theta\right)
\left[\chi''_{\kappa_-}(\omega)+\chi''_{\kappa_-}(-\omega)\right]\right)
\end{multline}
\end{widetext}
 From ~(\ref{ksi}), and ~(\ref{GammaT}) we obtain the voltage
amplitude $V_T$, and the phase, $\chi$: $V_T  = \omega \omega _T^2
L_T I_0 /\sqrt {\xi ^2  + \omega ^2 \Gamma _T^2 }$, and $tg\chi =
\xi /\omega \Gamma _T$.

These expressions have two different parts. The terms which are
proportional to $\varepsilon$ are due to transitions between
the Rabi levels. These terms have a resonance at the Rabi frequency and they
vanish at the optimal point (at $\varepsilon=0$). Other,
nonresonant, terms come from the transitions at the frequency
$\omega_0$ between neighboring manifolds. The contribution of these
terms does not vanish at the optimal point.

By assuming that the gap frequency $\omega_0$ is much more than $\omega$
and the rates $\Gamma$'s, we obtain at $\varepsilon=0$ with the
help of the susceptibilities (\ref{chikappa-approx}),
(\ref{chikappaapprox}), and (\ref{chikappa+approx}) the expressions
for low frequency detuning $\xi$ and friction $\Gamma_T$:
\begin{multline}\label{ksi_appr}
    \xi  = \omega _T^2  - \omega ^2-2\frac{{k^2L_q
    I_q^2}\omega^2_T}{{\Delta}}\widetilde{\rho}_+^{(0)}\\
    \times\left[1+\frac{\delta^2}{\Gamma_{\varphi}^2}-\frac{\delta}{\omega_0}
    \left(1+\frac{\Omega^2_1+\delta^2}{2\Gamma^2_{\varphi}}\right)\right]
\end{multline}
\begin{multline}\label{GammaT_appr}
    \Gamma _T  = \gamma _T+\frac{{k^2 L_qI_q^2}}{\Delta}
    \left(\frac{\delta}{\omega_0}\right)\Gamma_{\varphi} \widetilde{\rho}_+^{(0)}\\
    \times\left[3+\frac{2\Omega_R^2}{\Gamma_{\varphi}^2}+\frac{\Omega_R^2}{T_1\Gamma_{\varphi}^3}
    +\frac{\delta^2}{\Omega_R^2}\left(1+\frac{2\Omega_R^2}{\Gamma_{\varphi}^2}-\frac{\Omega_R^2}{T_1\Gamma_{\varphi}^3}\right)\right]
    \end{multline}
    where
\begin{equation}\label{dm+0}
    \widetilde{\rho}_+^{(0)}=\frac{\Gamma_{\varphi}^2\Gamma_-T_1}{\Gamma_{\varphi}^2+A_1\Omega^2_RT_1}
\end{equation}
At the point of resonance ($\delta=0$) the response of the flux
qubit is purely  inductive, $\Gamma_T=\gamma_T$. It is interesting
to note that expression (\ref{GammaT_appr}) predicts the
"cooling down" of the irradiated qubit due to its interaction with the tank circuit.
Since $\Gamma_-T_1$ is negative (see (\ref{balance1})) the
quantity $\Gamma_T$ at $\delta>0$ is less than $\gamma_T$, i. e. the quality factor of the tank is increased.
The energy from the qubit transfers to the tank: the qubit is cooled, the tank is heated. The
effect is more pronounced if $(\delta\gg\Omega_1)$. For this case
expression (\ref{GammaT_appr}) is reduced to:
\begin{equation}\label{GammaT reduced}
     \Gamma _T  = \gamma _T-4\frac{{k^2 L_qI_q^2}}{\Delta}
    \left(\frac{\delta}{\omega_0}\right)\Gamma_{\varphi}\tanh\left(\frac{\Delta}{2k_BT}\right)
\end{equation}
 Similar effects have been
observed in \cite{Ilich2}. These experiments showed an increase in
the quality of the tank resonator by approximately a factor of 2. For the
estimation of $\Gamma_T$ we take the parameters of the flux qubit
used in \cite{Ilich2}: $\Delta/\hbar=1$ GHz, $L_q=24$ pH,
$\omega_T/2\pi=6.284$ MHz, the tank quality factor $Q_T=1850$,
$\gamma_T=\omega_T/Q_T=2.0\times10^4$, $I_q=600$ nA,
$L_T=0.2\mu$H, $M=70$ pH, $\Gamma_{\varphi}=8\times10^5c^{-1}$, and
$\Omega_1\sim\omega_T$.
The result is
$\Gamma_T\approx\gamma_T-2\times10^4\left(\frac{\delta}{\omega_0}\right)$.
For $\delta\approx0.1\omega_0$ we obtain an increase in the
quality factor of the tank by approximately 15\%. Therefore, for
the parameters we used expression (\ref{GammaT_appr}) gives a
relatively weak effect.

As was shown by Ju. Hauss et al. (2007)\cite{Hauss1}, an additional contribution to
the heating of the tank, which can explain the observed amplitude,
appears in the second order terms in $G$ (which is beyond our linear
approximation) at twice the Rabi frequency $\omega\approx2\Omega_R$.
The effect appears at nonzero high frequency detuning $\delta$ as
a result of the "negative temperature" of the Rabi levels mentioned before
in connection with Eq.(\ref{st0}).
\subsection{Nuclear magnetic resonance}
Another application of the method can be made to nuclear magnetic resonance (NMR). The basic
input scheme of any NMR device is similar to that shown in Fig.
\ref{Fig3}, where  a sample with a substance under study replaces
the flux qubit loop. The common mode of operation
of the NMR device is to polarize the sample with a relatively high static
magnetic field  $B_0$ and at the same time apply a time dependent magnetic field
$B_1\cos\omega_0 t$ perpendicular to $B_0$. The amplitude
of the excitation signal is rather low ($B_1\ll B_0$) and it is tuned
to the NMR resonance frequency ($\omega_0\approx\gamma B_0$), where
$\gamma$ is the gyromagnetic ratio. The tank circuit detects the time
derivatives (Faraday law of induction) of the transversal
magnetizations $M_X$ and $M_Y$ which oscillate with the high NMR
frequency $\gamma B_0$. In addition, under the excitation
$B_1\cos\omega_0 t$ the longitudinal magnetization $M_Z$
oscillates with the Rabi frequency $\Omega_R\approx\gamma B_1$, and
sidebands $\omega_0\pm\Omega_R$ appear in the transversal components
$M_X$ and $M_Y$. However, due to coupling to the environment,
which is described by the relaxation $T_1$ and dephasing $T_2$ times in
Bloch equations, the Rabi oscillations undergo fast decay and
disappear from the output signal. The application to the sample of
another low frequency signal  $B_{LF}\cos\omega t$, which is tuned
to the Rabi frequency ($\omega\approx\Omega_R$) causes a
persistent low frequency oscillation of the magnetization with its
resonance at the Rabi frequency.

The case of NMR differs from the previous example in that here we deal
with a real spin-$1/2$ particle. Therefore, all components
$\langle\sigma_i\rangle$ ($i=X, Y, Z$) of the averaged spin operator
and their time derivatives are accessible for the measurements. As
distinct from the flux qubit case, where the low and high
frequency excitations are coupled only to $\sigma_z$ (see Eq.
\ref{Ham2}) in NMR we may couple the low frequency and high
frequency fields separately to any component of the spin operator. In
order to simplify the problem we consider the following
Hamiltonian for the spin-$1/2$ particle in an external magnetic
field:
\begin{multline}\label{Ham_nmr}
    H=\frac{\hbar\gamma
    B_0}{2}\sigma_Z+\frac{\hbar\gamma}{2}\sigma_ZB_{LF}\cos\omega
    t+\frac{\hbar\gamma}{2}\sigma_XB_1\cos\omega_0t
\end{multline}
where we choose the $z$ axis to be along the polarizing field $B_0$, and the sigh in (\ref{Ham_nmr}) is consistent with our convention $\sigma_Z|a\rangle=-|a\rangle$,
$\sigma_Z|b\rangle=|b\rangle$.

The high frequency excitation signal $B_1\cos\omega_0t$, which is
tuned to the NMR resonance frequency $\gamma B_0$, is applied along
the $x$ axis, and a low frequency signal $B_{LF}\cos\omega t$, which
is tuned to the Rabi frequency, $\omega\approx\Omega_R$, is applied
along the $z$ axis. It is just the low frequency component in
(\ref{Ham_nmr}) that causes the transitions between the Rabi levels.

Since the low frequency component is applied only along one axis
(as distinct from (\ref{HamLF})), $g_2=0$, and equations
(\ref{roLF})-(\ref{roLF3}) for the $\rho$'s are decoupled from
equations (\ref{kappa}), (\ref{kappa+}), and (\ref{kappa-}) for
the $\kappa$'s. This leads to a nonzero
solution for the $\kappa$'s only in the second order in low frequency
amplitude $g_1$. Therefore, in this case we have low frequency
responses only for the quantities $\langle\sigma_Z\rangle$ and
$\langle\frac{d\sigma_Z}{dt}\rangle$, which are given by
equations (\ref{sigma-z}) and (\ref{dszdt1}). In the equations for
the $\rho$'s and $\chi_{\rho}$'s we should substitute $\gamma
B_{LF}$ for $g_1$, and $\gamma$ for
$2\varepsilon/\hbar\Delta_{\varepsilon}$, respectively.

The macroscopic magnetic moment of a sample with $N$ spin-$1/2$
particles is as follows:
\begin{equation}\label{mz}
    M_Z=\frac{N\gamma\hbar}{2}\langle\sigma_Z\rangle
\end{equation}
The stationary magnetization $M_Z^{(ST)}$ is given by the stationary solution of (\ref{dszdt1}) or by the stationary part of (\ref{SgZ}). By making of the substitutions $\Gamma_{\varphi}=1/T_2$, $\Omega_1=\gamma B_1$, $\Delta_{\epsilon}=\hbar\gamma B_0$, we obtain for $M_Z^{(ST)}$ a well known expression \cite{Bloch1}:

\begin{equation}\label{mzst}
    M_Z^{(ST)}=M_0\frac{1+\left(T_2\delta\right)^2}{1+\left(T_2\delta\right)^2+T_1T_2\left(\gamma B_1\right)^2}
\end{equation}
where
\[
M_0=-\tanh\left(\frac{\hbar\gamma B_0}{2k_BT}\right)
\]

According to (\ref{sigma-z}) the low frequency response of $M_Z(t)$ is
\begin{equation}\label{mz-lf-resp}
    M_Z^{(LF)}(t)=\frac{N\gamma\hbar}{2}\left(\rho^{(1)}(t)\cos2\theta+\rho^{(1)}_+(t)\sin2\theta\right)
\end{equation}
where $\rho^{(1)}(t)$ and $\rho^{(1)}_+(t)$ can be expressed in
terms of the real ($\chi'$) and imaginary ($\chi''$) parts of their
corresponding susceptibilities:
$\rho^{(1)}(t)=B_{LF}\left(\chi'_{\rho}(\omega)\cos\omega
t-\chi''_{\rho}(\omega)\sin\omega t\right)$, and
$\rho^{(1)}_+(t)=B_{LF}\left(\chi'_{\rho_+}(\omega)\cos\omega
t-\chi''_{\rho_+}(\omega)\sin\omega t\right)$.

In the same way we obtain from (\ref{dszdt1}) the low frequency response for $\frac{dM_Z}{dt}$:
\begin{equation}\label{dmz-dt}
    \frac{dM_Z^{(LF)}}{dt}=
     -i\frac{N\gamma\hbar}{2}\Omega_R\sin2\theta\rho^{(1)}_-(t)-\frac{M_Z^{(LF)}(t)}{T_1}
\end{equation}
where
$\rho^{(1)}_-(t)=iB_{LF}\left(\chi'_{\rho_-}(\omega)\sin\omega
t+\chi''_{\rho_-}(\omega)\cos\omega t\right)$.

The corresponding susceptibilities are given by Eqs. (\ref{hiro}),
(\ref{hiro+}), and (\ref{hiro-}) or for small high frequency detuning
$(\delta\ll\Omega_R)$ by Eqs.(\ref{hi}), (\ref{hi+}), and (\ref{hi-})
where we should substitute $\gamma$ for
$2\varepsilon/\hbar\Delta_{\varepsilon}$.

The current state of the art allows one to detect low frequency
oscillations either of $M_Z$  with the help of superconducting
quantum interferernce devices (SQUIDs)\cite{Green4,McD} or of
$dM_Z/dt$ by a high quality resonant tank circuit\cite{Appelt}.
Below we write down the explicit form of $M_Z^{(LF)}(t)$ for the case of
small high frequency detuning ($\delta\ll\Omega_R$). In this case
the susceptibilities are as follows:

\begin{equation}\label{hinmr}
    \chi_{\rho}(\omega)=-\frac{\gamma\Omega_R}{\Gamma_{\varphi}}
    \frac{\rho_+^{(0)}}{i\omega+\Gamma_{\varphi}}
\end{equation}

\begin{equation}\label{hi+nmr}
    \chi_{\rho_+}(\omega)=-\delta\gamma
    \frac{\rho_+^{(0)}}{\Gamma_{\varphi}d(\omega)}
    \left[i\omega+\frac{1}{T_1}-\frac{\Omega_R^2}{\Gamma_{\varphi}}\right]
\end{equation}

\begin{multline}\label{hi-nmr}
    \chi_{\rho_-}(\omega)=i\delta\gamma
    \frac{\Omega_R\rho_+^{(0)}}{\Gamma_{\varphi}^2
    d(\omega)}
\frac{\left(i\omega+\frac{1}{T_1}\right)\left(i\omega+2\Gamma_{\varphi}\right)
}{\left(i\omega+\Gamma_{\varphi}\right)}
\end{multline}
where $d(\omega)$ is defined in (\ref{dom}), $\Omega_R=\gamma
B_1$, and
\begin{equation}\label{ronmr}
    \rho_+^{(0)}=\frac{\Gamma_{\varphi}}{\Gamma_{\varphi}+\gamma^2B_1^2T_1}
    \tanh\left(\frac{\hbar\gamma B_0}{2k_BT}\right)
\end{equation}
For $M_Z^{(LF)}(t)$ we obtain:

\begin{multline}\label{mzlf}
M_Z^{(LF)}(t)  = \frac{{N\hbar \gamma }}{2}\left( {\gamma B_{LF} }
\right)\delta \frac{{\Omega _R^2 }}{{\Gamma _\varphi ^2 }}\rho _ +
^{(0)}\times \\\left[ {A_Z(\omega )\cos \omega t + B_Z(\omega )\sin \omega
t} \right]
\end{multline}

where
\begin{equation}\label{aomega}
A_Z(\omega ) = \frac{{\left[ {\omega ^4  - \omega ^2 \left(
{\Omega _R^2  - 3\Gamma _\varphi ^2 } \right) - 2\Gamma _\varphi
^2 \left( {\Omega _R^2  + \frac{{\Gamma _\varphi  }}{{T_1 }}}
\right)} \right]}}{{\left( {\omega ^2  + \Gamma _\varphi ^2 }
\right)\left[ {\left( {\widetilde\Omega _R^2  - \omega ^2 }
\right)^2  + \omega ^2 \left( {\Gamma _\varphi   + T_1^{ - 1} }
\right)^2 } \right]}}
\end{equation}
\begin{equation}\label{bomega}
B_Z(\omega ) = \frac{\omega }{{T_1 }}\frac{{\left[ {\omega ^2  +
\Gamma _\varphi  T_1 \left( {\Omega _R^2  + \frac{{\Gamma _\varphi
}}{{T_1 }} + 2\Gamma _\varphi ^2 } \right)} \right]}}{{\left(
{\omega ^2  + \Gamma _\varphi ^2 } \right)\left[ {\left(
{\widetilde\Omega _R^2  - \omega ^2 } \right)^2  + \omega ^2
\left( {\Gamma _\varphi   + T_1^{ - 1} } \right)^2 } \right]}}
\end{equation}

$\widetilde{\Omega}_R^2=\Omega_R^2+\Gamma_{\varphi}/T_1$.

\section{Conclusion}
In this paper in the frame of the dressed state approach we have
analyzed the interaction of a dissipative two level quantum system
with high and low frequency excitations. We have found a linear
response of the coupled TLS and associated photon field system to a signal whose
frequency is of the order of the Rabi frequency. The response of
the system exhibits an undamped low frequency oscillation, whose
amplitude has a clear resonance at the Rabi frequency with the
width being dependent on the damping rates of the system.
The explicit expressions for low frequency susceptibility of the TLS,
which we obtained in the paper, have a rather general nature and can be applied for the investigation of specific physical systems.
As an illustration of our approach we study in detail two rather different physical systems:
the flux qubit and real spin under NMR conditions.

\begin{acknowledgments}
The author thanks  Evgeni Il'ichev for many stimulating
discussions.  The financial support from the ESF under grant No.
1030 and DFG under grant IL 150/1-1 as well as the hospitality of IPHT (Jena, Germany) is greatly
acknowledged.
\end{acknowledgments}
\begin{widetext}
\begin{center}
{\textbf{Appendix}}
\end{center}
\subsection {Calculation of matrix elements of $\widehat{L}$ in the
dressed state basis}

With the aid of (\ref{WF+}), (\ref{WF-}) and (\ref{L1}), we obtain
for $\langle 1,N|\widehat{L}|1,N\rangle$:

\begin{multline}\label{L4}
\langle 1,N|\widehat{L}|1,N\rangle=\\\sin^2\theta\langle
a,N|\widehat{L}|a,N\rangle+\cos^2\theta\langle
b,N-1|\widehat{L}|b,N-1\rangle+\sin\theta\cos\theta\left[\langle
a,N|\widehat{L}|b,N-1\rangle+\langle b,N-1|\widehat{L}|a,N\rangle\right]=\\
\sin^2\theta\left[-\Gamma_{\uparrow}\langle
a,N|\sigma|a,N\rangle+\Gamma_{\downarrow}\langle
b,N|\sigma|b,N\rangle\right]+
\cos^2\theta\left[-\Gamma_{\downarrow}\langle
b,N-1|\sigma|b,N-1\rangle+\Gamma_{\uparrow}\langle
a,N-1|\sigma|a,N-1\rangle\right]-\\
\Gamma_{\varphi}\sin\theta\cos\theta\left[\langle
a,N|\sigma|b,N-1\rangle+\langle b,N-1|\sigma|a,N\rangle\right]
\end{multline}
Further transformation requires the substitution of the uncoupled
states in (\ref{L4}) with the dressed states by using Eqs.
(\ref{WFa}), and (\ref{WFb}). As a result we obtain:
\begin{multline}\label{L5}
\langle 1,N|\widehat{L}|1,N\rangle=\\-\langle
1,N|\sigma|1,N\rangle\left[\Gamma_{\uparrow}\sin^4\theta+\Gamma_{\downarrow}\cos^4\theta+
2\Gamma_{\varphi}\sin^2\theta\cos^2\theta\right]-\langle
2,N|\sigma|2,N\rangle\sin^2\theta\cos^2\theta\left[\Gamma_{\uparrow}
+\Gamma_{\downarrow}-2\Gamma_{\varphi}\right]+\\
\left[\langle 1,N|\sigma|2,N\rangle+\langle
2,N|\sigma|1,N\rangle\right]\sin\theta\cos\theta\left[\Gamma_{\uparrow}\sin^2\theta
-\Gamma_{\downarrow}\cos^2\theta+\Gamma_{\varphi}\cos2\theta\right]+
\sin^2\theta\cos^2\theta\Gamma_{\downarrow}\langle
1,N+1|\sigma|1,N+1\rangle+\\\sin^2\theta\cos^2\theta\Gamma_{\uparrow}\langle
1,N-1|\sigma|1,N-1\rangle+\sin^4\theta\Gamma_{\downarrow}\langle
2,N+1|\sigma|2,N+1\rangle+\cos^4\theta\Gamma_{\uparrow}\langle
2,N-1|\sigma|2,N-1\rangle+\\\left[\langle
1,N+1|\sigma|2,N+1\rangle+\langle
2,N+1|\sigma|1,N+1\rangle\right]\Gamma_{\downarrow}\sin^3\theta\cos\theta-\\
\left[\langle 1,N-1|\sigma|2,N-1\rangle+\langle
2,N-1|\sigma|1,N-1\rangle\right]\Gamma_{\uparrow}\sin\theta\cos^3\theta
\end{multline}
We do not explicitly write here the other matrix elements of
$\widehat L$, such as $\langle 2,N|\widehat{L}|2,N\rangle$, $\langle 1,N|\widehat L|2,N\rangle$,
 $\langle {1,N} |\hat L| {1,N - 1} \rangle$, $\langle {2,N} |\hat L| {2,N - 1} \rangle$,
 $\langle {1,N} |\hat L| {2,N - 1} \rangle $, $\langle {2,N} |\hat L| {1,N - 1} \rangle$, which can be obtained by similar procedure.

\subsection{Calculations of the averages of Pauli spin operators
$\langle\sigma_X\rangle$, $\langle\sigma_Y\rangle$, and
$\langle\sigma_Z\rangle$ in the dressed state basis}

\begin{multline}\label{Trsigmaz1}
    \langle\sigma_Z\rangle=Tr\left(\sigma\sigma_Z\right)=
    \sum_N\langle1,N|\sigma\sigma_Z|1,N\rangle+\sum_N\langle
    2,N|\sigma\sigma_Z|2,N\rangle=
\sum_N\langle1,N|\sigma|1,N\rangle\langle 1,N|\sigma_Z|1,N\rangle+\\
    \sum_N\langle1,N|\sigma|2,N\rangle\langle2,N|\sigma_Z|1,N\rangle+
    \sum_N\langle2,N|\sigma|1,N\rangle\langle1,N|\sigma_Z|2,N\rangle+
\sum_N\langle2,N|\sigma|2,N\rangle\langle2,N|\sigma_Z|2,N\rangle
\end{multline}
\begin{multline}\label{Trsigmax}
    \langle\sigma_i\rangle=Tr\left(\sigma\sigma_i\right)=
    \sum_N\langle1,N|\sigma\sigma_i|1,N\rangle+\sum_N\langle 2,N|\sigma\sigma_i|2,N\rangle=\\
\sum_N\langle1,N|\sigma|1,N+1\rangle\langle 1,N+1|\sigma_i|1,N\rangle+\sum_N\langle1,N|\sigma|1,N-1\rangle\langle 1,N-1|\sigma_i|1,N\rangle+\\
    \sum_N\langle1,N|\sigma|2,N+1\rangle\langle2,N+1|\sigma_i|1,N\rangle+\sum_N\langle1,N|\sigma|2,N-1\rangle\langle2,N-1|\sigma_i|1,N\rangle+\\
    \sum_N\langle2,N|\sigma|1,N+1\rangle\langle1,N+1|\sigma_i|2,N\rangle+\sum_N\langle2,N|\sigma|1,N-1\rangle\langle1,N-1|\sigma_i|2,N\rangle+\\
\sum_N\langle2,N|\sigma|2,N+1\rangle\langle2,N+1|\sigma_i|2,N\rangle+\sum_N\langle2,N|\sigma|2,N+1\rangle\langle2,N+1|\sigma_i|2,N\rangle
\end{multline}

where $i=X,Y$.

\end{widetext}


\begin{thebibliography}{99}

\bibitem{Rabi} I. I. Rabi, Phys. Rev. \textbf{51}, 652 (1937).

\bibitem{beam} Atomic and Molecular Beams: The State of The Art
2000, (Roger Compargue, ed.) Springer Verlag Telos, 2001.

\bibitem{Raimond} J. M. Raimond, M. Brune, and S. Haroche, Rev. Mod. Phys. \textbf{73}, 565 (2001).

\bibitem{Makh} Makhlin Y. Sch\"on G. and Shnirman A., Rev. Mod. Phys., 73
(2001) 357.

\bibitem{Averin} D. V. Averin, in: "\emph{Exploring the quantum/classical frontier:
recent advances in macroscopic quantum phenomena}", Ed. by J.R.
Friedman and S. Han, (Nova Publishes, Hauppauge, NY, 2002), p.
441; cond-mat/0004364.

\bibitem{Kor1} A.N. Korotkov and D.V. Averin, Phys. Rev. B \textbf{64}, 165310 (2001).

\bibitem{Kor2} A.N. Korotkov, Phys. Rev. B \textbf{63}, 115403 (2001).

\bibitem{Nakamura} Y. Nakamura, Yu. A. Pashkin, and  J. S. Tsai,
Phys. Rev. Lett. \textbf{87}, 246601 (2001).

\bibitem{Vion} D. Vion, A. Aassime, A. Cottet, P. Joyez, H. Pothier,
C. Urbina, D. Esteve, M. H. Devoret, Science \textbf{296}, 886
(2002).

\bibitem{Martinis} J.M. Martinis, S. Nam, J. Aumentado, C. Urbina,  Phys. Rev. Lett. \textbf{89}, 117901 (2002).%S. Nam, J. Aumentado, and C. Urbina,

\bibitem{Chiorescu} I. Chiorescu, Y. Nakamura, C. J. P. M. Harmans, and J. E. Mooij, Science \textbf{299}, 1869 (2003).

\bibitem{Ilich1}E. Il'ichev, A.Yu. Smirnov, M. Grajcar, A. Izmalkov, D. Born,
N. Oukhanski, Th. Wagner, W. Krech, H.-G. Meyer, and A. Zagoskin,
Fizika Nizkikh Temperatur, \textbf{30}, 823 (2004).

\bibitem{Ilich2} E. Il'ichev, N. Oukhanski, A. Izmalkov, Th. Wagner, M. Grajcar,
H.-G. Meyer, A.Yu. Smirnov, A. Maassen van den Brink, M.H.S. Amin,
A.M. Zagoskin, Phys. Rev. Lett. 91, 097906 (2003).

\bibitem{Smirnov1} A. Yu. Smirnov, Phys. Rev. B \textbf{68}, 134514 (2003).

\bibitem{Smirnov2} A. Yu. Smirnov, e-print archive cond-mat/0306004.

\bibitem{Hauss} Ju. Hauss, Rabi Spektroskopie an Qubit-Oszillator
Systemen, Diploma thesis, Karlsruhe University, 2006.

\bibitem{Hauss1}Ju. Hauss, A. Fedorov, C. Hutter, A. Shnirman, and G.
Sch\"on, e-print archive cond-mat/0701041.

\bibitem{Green1} Ya. S. Greenberg and E. Il'ichev, e-print archive quant-ph/0502187.

\bibitem{Green2} Ya. S. Greenberg, E. Il'ichev and A. Izmalkov,
Europhys. Lett., \textbf{72}, 880 (2005)

\bibitem{Coen} C. Coen-Tannoudji, J. Dupont-Rock, G. Grynberg,
Atom-Photon Interactions. Basic Prinsiples and Applications. (John
Wiley and Sons, 1998).

\bibitem{Grif} Grifoni M. and H¨anggi P., Phys. Rep.,
304 (1998) 229.

\bibitem{com1} Frequently, the term "Rabi frequency" is associated
with the quantity $\Omega_1$. Here we call by this term the
quantity $\Omega_R$, the frequency with which the population
oscillates if the high frequency detuning $\delta$ is different
from zero.

\bibitem{Bloch} F. Bloch, Phys. Rev.\textbf{105}, 1206 (1957).
\bibitem{Red} A. G. Redfield, IBM J. Res. Dev. \textbf{1}, 19 (1957).
\bibitem{com} It is true that if the system is subjected to a strong external driving the $\Gamma$'s rates and the corresponding rate equations can be substantially modified \cite{Smirnov1,Hartmann}. However, even in this case the most of experiments can adequately be explained  by the Bloch equations with phenomenological $\Gamma$'s rates as fitting parameters (see, for example,\cite{Saito}).
\bibitem{Hartmann} L. Hartmann, I. Goychuk, M. Grifoni, P. Hanggi,  Phys. Rev. E \textbf{61}, R4687 (2000).

\bibitem{Saito} S. Saito, M. Thorwart, H. Tanaka, M. Ueda, H. Nakano, K. Semba, and H. Takayanagi, Phys. Rev. Lett. \textbf{93}, 037001 (2004).

\bibitem{Mooij} J. E. Mooij, T. P. Orlando, L. Levitov, L. Tian, C. H. van der Wal,
and S. Lloyd, Science \textbf{285}, 1036 (1999).

\bibitem{Orlando} T.P. Orlando, J.E. Mooij, L. Tian, C.H. van der Wal, L.~Levitov,
S. Lloyd, and J.J. Mazo, Phys. Rev. B {\bf 60}, 15398 (1999).

\bibitem{Green3} Ya. S. Greenberg, A. Izmalkov, M. Grajcar, E. Il'ichev,
 W. Krech, H.-G. Meyer, M. H. S. Amin,
and A. Maassen van den Brink, Phys. Rev B \textbf{66}, 214525
(2002).


\bibitem{Grajcar} M. Grajcar, A. Izmalkov, E. Il'ichev, Th. Wagner, N. Oukhanski,
 U. Hu¨bner, T. May, I. Zhilyaev, H. E. Hoenig,
Ya. S. Greenberg, V. I. Shnyrkov, D. Born, W. Krech, H.-G. Meyer,
A. Maassen van den Brink, and M. H. S. Amin, Phys. Rev. B
\textbf{69}, 060501(R) (2004).
\bibitem{Bloch1}  F. Bloch, Phys. Rev.\textbf{70}, 460 (1946).
\bibitem{Green4} Ya. S. Greenberg Rev. Mod. Phys. \textbf{70}, 175 (1998).

\bibitem{McD} R. McDermot, A. H. Trabesinger, M. M\"uck, E. L.
Hahn, A. Pines, and J. Clarke, Science \textbf{295}, 2247 (2002).

\bibitem{Appelt} S. Appelt, F. W. H\"asing, H. K\"uhn, J. Perlo,
and B. Bl\"umich, Phys. Rev. Lett. \textbf{94}, 197602 (2005).





\end{thebibliography}
\end{document}